\documentclass[10pt,aps,preprintnumbers, prd, twocolumn, showpacs, floatfix,
letterpaper, superscriptaddress,nofootinbib]{revtex4-1}
\pdfoutput=1

\usepackage{graphicx,epsfig}
\usepackage{amsmath,mathrsfs,amsfonts}
\usepackage{amssymb}
\usepackage{longtable}
\usepackage{multirow}
\usepackage{enumerate}
\usepackage{dcolumn}
\usepackage{bm}
\usepackage{subfigure}
\usepackage{color}
\usepackage{relsize}

\newcommand{\be}{\begin{equation}}
\newcommand{\ee}{\end{equation}}
\newcommand{\bea}{\begin{eqnarray}}
\newcommand{\eea}{\end{eqnarray}}

\begin{document}

\title{Gauge-invariant non-universal scalar couplings in teleparallel gravity}

\author{Giovanni Otalora}
\email{giovanni.otalora@academicos.uta.cl}
\affiliation{Departamento de F\'isica, Facultad de Ciencias, Universidad de Tarapac\'a, Casilla 7-D, Arica, Chile}

\author{Rudranil Roy}
\email{rudranil.roy@alumnos.uta.cl}
\affiliation{Departamento de F\'isica, Facultad de Ciencias, Universidad de Tarapac\'a, Casilla 7-D, Arica, Chile}
\affiliation{Departamento de F\'{\i}sica, Facultad de Ciencias, Universidad de La Serena, Avenida Cisternas 1200, La Serena, Chile}
\affiliation{Instituto de F\'{\i}sica y Astronom\'{\i}a, Universidad de Valpara\'{\i}so, Gran Breta\~{n}a 1111, Valpara\'{\i}so, Chile}

\date{\today}

\begin{abstract}
We develop a gauge-invariant teleparallel framework in which the universality of the gravitational coupling is relaxed for a scalar field, while the gravitational sector retains the form of the teleparallel equivalent of General Relativity. The non-universal interaction is encoded in a scalar-sector tetrad whose coupling to the universal translational gauge potential is controlled by a field-dependent gravitational charge. A St\"uckelberg completion preserves local translational gauge invariance, and we derive the gravitational, scalar-field, and St\"uckelberg equations, together with the antisymmetric constraint and their consistency relations. As a first application, we investigate homogeneous and isotropic cosmology, where the St\"uckelberg constraint provides a first integral governing the evolution of the non-universal
coupling. In the kination-dominated regime, the standard
stiff-fluid behavior is recovered asymptotically, whereas in
the potential-dominated regime, accelerated expansion requires
additional conditions beyond potential domination. We identify
compatible expanding solutions in which departures from universality are dynamically suppressed, including examples approaching power-law inflation or asymptotically de Sitter expansion. Including a universally coupled perfect fluid and a constant scalar potential, we obtain a late-time solution exhibiting a subdominant matter-like contribution to the background Friedmann equation and an asymptotic de Sitter limit. These results provide a gauge-theoretic realization of
non-universal scalar couplings and highlight their potential
implications for early- and late-time cosmology.
\end{abstract}


\maketitle

\section{Introduction}
\label{sec:introduction}
Scalar fields play a central role in gravitational physics and cosmology, appearing as inflatons \cite{Starobinsky:1980te,Guth:1980zm,Linde:1981mu,Albrecht:1982wi,Lyth:1998xn}, dark-energy candidates \cite{Copeland:2006wr}, scalar field dark-matter candidates \cite{Hu:2000ke,Hui:2016ltb}, and as additional degrees of freedom in extensions of General Relativity (GR) \cite{Clifton:2011jh,Capozziello:2011et}. Their coupling to gravity is usually assumed to be universal, in accordance with the universality of free fall. This assumption also underlies the geometrical description of gravitation. If distinct matter sectors couple to gravity with different strengths, their motion cannot, in general, be simultaneously described by the geodesics of a single spacetime metric.

Teleparallel gravity provides a conceptually different perspective. In the teleparallel equivalent of General Relativity (TEGR), gravitation is formulated
as a gauge theory of spacetime translations. The fundamental gravitational variable is the translational gauge potential, torsion represents its field strength, and the metric is a derived quantity
\cite{Early-papers5,Early-papers6,Aldrovandi-Pereira-book,Pereira:2019woq}. For universal matter couplings,
TEGR is dynamically equivalent to GR, but its gauge interpretation becomes
particularly relevant once universality is relaxed. Aldrovandi, Pereira, and Vu
\cite{Aldrovandi:2003pa,Aldrovandi:2004fy} showed that the teleparallel description of a spinless
particle remains consistent when the gravitational and inertial masses differ. The translational gauge potential remains universal, while the dynamics depends
on the ratio between the gravitational and inertial masses, which plays the role
of a dimensionless gravitational charge. Non-universality can therefore reside
in the matter coupling without requiring a modification of the fundamental
gravitational field.

Scalar-dependent interactions arise naturally in string-inspired
extensions of GR, where dilaton couplings to matter may lead to
composition-dependent interactions and violations of the
equivalence principle. Damour and Polyakov
\cite{Damour:1994zq} showed that, provided the relevant
dilaton coupling functions share a common extremum,
cosmological evolution can drive the dilaton toward a
least-coupling point, suppressing its interactions with matter.
In the runaway-dilaton scenario of Damour, Piazza, and
Veneziano \cite{Damour:2002mi}, inflationary evolution instead
drives the dilaton toward infinite bare string coupling,
where its effective couplings to ordinary matter vanish
asymptotically. Unlike these scalar-mediated mechanisms,
the present construction introduces non-universality through
a scalar-dependent coupling to the teleparallel translational
gauge potential, such that the scalar field couples to
gravity through an effective tetrad distinct from the
universal one. Nevertheless, these works provide important
precedents for the cosmological suppression of non-universal
interactions.

More generally, additional scalar degrees of freedom constitute one of the most extensively studied sectors of modified gravity \cite{Clifton:2011jh,Capozziello:2011et}. Familiar curvature-based examples include scalar–tensor theories with nonminimal couplings of the form $F(\phi)R$ \cite{Fujii:2003pa,Faraoni:2004pi,Roy:2026vwx}, as well as $f(R)$ gravity, which can be represented in terms of an additional scalar degree of freedom \cite{DeFelice:2010aj,Burton-Villalobos:2025tgc,Leyva:2022zhz}. Analogous extensions have been widely developed in the teleparallel framework \cite{Cai:2015emx,Bahamonde:2021gfp}, including nonminimal scalar–torsion interactions of the form $F(\phi)T$ \cite{Geng:2011aj} and the more general $f(T,\phi)$ gravity theories \cite{Hohmann:2018rwf,Hohmann:2018vle}. Their cosmological applications include inflationary dynamics and primordial perturbations \cite{Gonzalez-Espinoza:2019ajd,GonzalezEspinoza:2020azh}, reconstruction of inflationary scenarios \cite{Gonzalez-Espinoza:2021qnv}, observational constraints on inflation \cite{Kumar:2025apf}, primordial black-hole production \cite{Villalobos-Silva:2026deu}, and late-time cosmology and dark-energy dynamics \cite{Gonzalez-Espinoza:2020jss,Gonzalez-Espinoza:2021mwr,Gonzalez-Espinoza:2022hui,Gonzalez-Espinoza:2023whd,Rodriguez-Benites:2024pce}.

The mechanism considered here is conceptually different. Motivated by
Ref.~\cite{Aldrovandi:2003pa}, we extend the notion of a non-universal
gravitational charge from the point-particle description to a scalar field by
allowing its coupling to the translational gauge potential to depend on the
field itself through a dimensionless function $\alpha(\phi)$. The gravitational
action remains that of TEGR and is constructed from a universal tetrad, while
the scalar field propagates according to a scalar-sector tetrad that couples to
the same translational gauge potential with a field-dependent strength. Thus,
unlike conventional nonminimal scalar--curvature or scalar--torsion theories,
in which the scalar field couples directly to curvature or torsion invariants
in the gravitational action, here the pure gravitational sector retains the
TEGR form with a constant gravitational coupling. The scalar--gravity
interaction instead originates from the field-dependent coupling of the scalar
sector to the universal translational gauge potential through its effective
tetrad. The universally coupled canonical scalar theory is continuously
recovered in the limit $\alpha(\phi)\rightarrow0$.

A field-dependent coupling to the translational gauge potential requires
particular care because the gauge potential transforms inhomogeneously under
local translations. Following the St\"uckelberg procedure
\cite{Ruegg:2003ps,Hinterbichler:2011tt}, we introduce four compensating fields
and combine them with the gauge potential to form a translationally invariant
quantity. This allows us to define the scalar-sector effective tetrad while
preserving local translational gauge invariance. The original form of the
scalar-sector tetrad is recovered in unitary gauge. The restored gauge symmetry
leads, through Noether's second theorem, to an off-shell identity relating the
corresponding field equations and provides a nontrivial consistency condition
for the gauge-invariant formulation. Furthermore, the effective
energy--momentum tensor sourcing the universal gravitational equations is
covariantly conserved, as required by the gravitational Bianchi identity.
Together, these properties establish the consistency of the gauge-invariant
non-universal construction.

The simultaneous presence of a universal and a scalar-sector tetrad should
also be distinguished from bimetric and bigravity theories
\cite{Hassan:2011zd,Hassan:2012wr}. In ghost-free bigravity, the two metrics,
or equivalently two tetrads, represent independent dynamical gravitational
fields with their own kinetic sectors and mutual interactions, giving rise to
massless and massive spin-2 modes. In the present construction, there is only
one fundamental gravitational tetrad governed by the TEGR action. The
scalar-sector tetrad is instead a derived quantity encoding the non-universal
coupling of the scalar field to the same translational gauge potential and
therefore does not constitute a second independent spin-2 gravitational field.

Related effective geometries can also emerge through different physical
mechanisms, including conformal and disformal transformations in teleparallel
gravity \cite{Golovnev:2019kcf,Hohmann:2019gmt}, semiclassical tetrad
corrections \cite{Chen:2021oal}, energy-dependent geometries in gravity's
rainbow \cite{Magueijo:2002xx,Leyva:2022zhz,Leyva:2021fuo}, and the effective
spin-dependent geometry induced by spin-gravity coupling
\cite{Otalora:2022ehd}. In the present theory, however, the scalar-sector
tetrad is not introduced as a transformation of the gravitational variables
or as a probe-dependent geometry, but arises from the field-dependent coupling
of the scalar sector to the same translational gauge potential that defines
the universal gravitational tetrad. The distinctive feature of the
construction is therefore not the occurrence of an effective scalar-dependent
geometry itself, but its origin in a relaxation of the universality of the
coupling to the translational gauge potential. We formulate the corresponding
gauge-invariant action and derive the complete field equations without
specifying a spacetime background. The gravitational equations are sourced by
an effective scalar energy-momentum tensor determined by the relation between
the universal and scalar-sector tetrads and are supplemented by an
antisymmetric constraint.

As a cosmological application, we investigate the homogeneous and isotropic
dynamics in a spatially flat Friedmann--Lema\^itre--Robertson--Walker (FLRW)
background. Using the proper diagonal tetrad for spatially flat FLRW geometry
\cite{Krssak:2015oua}, the translational gauge potential has only spatial
components, so that the non-universal coupling modifies the scalar-sector
scale factor while leaving the lapse unchanged. We derive the corresponding
Friedmann and scalar-field equations together with the St\"uckelberg
consistency condition and analyze the kination- and potential-dominated
regimes. The non-universal coupling modifies the effective equation of state
and can lead to background dynamics that differ from those of a universally
coupled canonical scalar field. In this configuration, the St\"uckelberg
equation provides a first integral that constrains the evolution of the
non-universal coupling.

We further include a universally coupled perfect fluid and consider an
approximately constant scalar potential to explore the late-time cosmological
behavior. For solutions continuously connected to the universal limit, the
non-universal coupling is dynamically suppressed as the Universe expands and
generates an effective contribution to the Friedmann equation scaling as
$a^{-3}$. The resulting background approaches a matter-to-de Sitter evolution,
with a matter-like contribution arising from the non-universal scalar sector
rather than from an additional universally coupled matter component. These
results show that the proposed non-universal coupling can modify the
cosmological dynamics while allowing the system to dynamically approach the
universal regime.

The paper is organized as follows. Section~\ref{sec:TG} reviews the gauge
formulation of teleparallel gravity, while Sec.~\ref{sec:nonuniversal_teleparallel} summarizes
its extension to non-universal gravitational couplings. In
Sec.~\ref{sec:scalar_nonuniversal} we introduce the gauge-invariant theory of a non-universally coupled scalar field, and in Sec.~\ref{sec:field_equations} we derive its field equations and consistency
conditions. Section~\ref{sec:cosmology} is devoted to homogeneous and isotropic
cosmology, including the kination-, potential-dominated, and late-time regimes.
Finally, we summarize our results in Sec.~\ref{sec:conclusions}.

\section{Teleparallel gravity}
\label{sec:TG}

Teleparallel gravity formulates gravitation as a gauge theory of local translations in the tangent-space fibers attached to each spacetime point~\cite{Early-papers5,Early-papers6,Aldrovandi-Pereira-book,Pereira:2019woq}. We denote spacetime coordinates by $x^\mu$ and tangent-space coordinates by $x^A$. In an orthonormal frame, the Minkowski metric on each tangent space has components $\eta_{AB}=\mathrm{diag}(-1,1,1,1)$.

The gravitational field is represented by $B^A{}_\mu$, the gauge potential associated with these local translations. In the Lorentz-covariant formulation, it enters the tetrad through
\begin{equation}
h^A{}_\mu=D_\mu x^A+B^A{}_\mu,
\qquad
D_\mu x^A=\partial_\mu x^A+\omega^A{}_{B\mu}x^B,
\label{eq:tetrad_TG}
\end{equation}
where $\omega^A{}_{B\mu}$ is the inertial spin connection. The tetrad then relates the tangent-space Minkowski metric to the spacetime metric and determines the corresponding volume density,
\begin{equation}
g_{\mu\nu}=\eta_{AB}h^A{}_\mu h^B{}_\nu,
\qquad
h\equiv\det(h^A{}_\mu)=\sqrt{-g}.
\label{eq:metric_TG}
\end{equation}

In addition to spacetime diffeomorphisms, the theory admits local Lorentz transformations of the tangent frame. Under a change of frame, the tetrad transforms as
\begin{equation}
h^A{}_\mu\rightarrow h^{\prime A}{}_\mu
=
\Lambda^A{}_B(x)h^B{}_\mu,
\qquad
\eta_{CD}\Lambda^C{}_A\Lambda^D{}_B=\eta_{AB}.
\label{eq:local_Lorentz}
\end{equation}
The spacetime metric is invariant under this transformation. The corresponding transformation of the inertial spin connection is
\begin{equation}
\omega^{\prime A}{}_{B\mu}
=
\Lambda^A{}_C\,
\omega^C{}_{D\mu}\,
(\Lambda^{-1})^D{}_B
+
\Lambda^A{}_C
\partial_\mu(\Lambda^{-1})^C{}_B .
\label{eq:spin_connection_transformation}
\end{equation}
In teleparallel gravity, the spin connection represents inertial effects and
locally takes the pure-gauge form
\cite{AldrovandiPereiraBook,Krssak:2015oua}
\begin{equation}
\omega^A{}_{B\mu}
=
\Lambda^A{}_C\,\partial_\mu\Lambda_B{}^C .
\label{eq:inertial_spin_connection}
\end{equation}
It is therefore curvature-free,
\begin{equation}
R^A{}_{B\mu\nu}(\omega)=0.
\label{eq:flat_spin_connection}
\end{equation}
A frame in which $\omega^A{}_{B\mu}=0$ defines the Weitzenb\"ock gauge.

Local translations act in the tangent-space fiber at each spacetime point. Under the transformations
\begin{equation}
x^A\rightarrow x^A+\epsilon^A(x),
\qquad
B^A{}_\mu\rightarrow B^A{}_\mu-D_\mu\epsilon^A,
\label{eq:local_translation}
\end{equation}
the tetrad is invariant,
$\delta_\epsilon h^A{}_\mu=0$. Unlike ordinary internal gauge theories, the tangent space is soldered to
spacetime through the tetrad, establishing a local correspondence between
tangent-space and spacetime directions. Local translations are therefore
closely related to spacetime transformations, while representing the internal
gauge symmetry associated with the translational gravitational potential
$B^A{}_\mu$. Diffeomorphisms, local Lorentz transformations, and local
translations thus have distinct geometric roles.

The teleparallel affine connection is
\begin{equation}
\Gamma^\rho{}_{\nu\mu}
=
h_A{}^\rho
\left(
\partial_\mu h^A{}_\nu
+
\omega^A{}_{B\mu}h^B{}_\nu
\right),
\label{eq:teleparallel_connection}
\end{equation}
and the corresponding torsion tensor is
\begin{align}
T^\rho{}_{\mu\nu}
&=
\Gamma^\rho{}_{\nu\mu}
-
\Gamma^\rho{}_{\mu\nu}
\nonumber\\
&=
h_A{}^\rho
\left(
\partial_\mu h^A{}_\nu
-\partial_\nu h^A{}_\mu
+\omega^A{}_{B\mu}h^B{}_\nu
-\omega^A{}_{B\nu}h^B{}_\mu
\right).
\label{eq:torsion_tensor}
\end{align}
Equivalently,
\begin{equation}
T^\rho{}_{\mu\nu}=h_A{}^\rho T^A{}_{\mu\nu},
\label{eq:torsion_relation}
\end{equation}
where
\begin{equation}
T^A{}_{\mu\nu}
=
D_\mu h^A{}_\nu-D_\nu h^A{}_\mu .
\label{eq:tangent_torsion}
\end{equation}
Using the decomposition \eqref{eq:tetrad_TG}, the torsion becomes
\begin{equation}
T^A{}_{\mu\nu}
=
[D_\mu,D_\nu]x^A
+
D_\mu B^A{}_\nu-D_\nu B^A{}_\mu .
\end{equation}
Since the inertial spin connection is curvature-free,
\begin{equation}
[D_\mu,D_\nu]x^A
=
R^A{}_{B\mu\nu}(\omega)x^B=0,
\end{equation}
and hence
\begin{equation}
T^A{}_{\mu\nu}
=
D_\mu B^A{}_\nu-D_\nu B^A{}_\mu .
\label{eq:translation_field_strength_covariant}
\end{equation}
Thus, the gravitational gauge degrees of freedom are encoded in the translational gauge potential $B^A{}_\mu$, whose associated field strength is represented by the torsion $T^A{}_{\mu\nu}$.

The contortion tensor and superpotential are defined, respectively, by
\begin{align}
K^{\mu\nu}{}_{\rho}
&=
-\frac{1}{2}
\left(
T^{\mu\nu}{}_{\rho}
-
T^{\nu\mu}{}_{\rho}
-
T_{\rho}{}^{\mu\nu}
\right),
\\
S_{\rho}{}^{\mu\nu}
&=
\frac{1}{2}
\left[
K^{\mu\nu}{}_{\rho}
+
\delta^\mu_{\rho}T^{\alpha\nu}{}_{\alpha}
-
\delta^\nu_{\rho}T^{\alpha\mu}{}_{\alpha}
\right].
\label{eq:superpotential}
\end{align}
The overall factor $1/2$ in the definition of
$S_{\rho}{}^{\mu\nu}$ fixes the convention adopted throughout this work. The
torsion scalar is
\begin{equation}
T=S_{\rho}{}^{\mu\nu}T^\rho{}_{\mu\nu},
\label{eq:torsion_scalar}
\end{equation}
or, equivalently,
\begin{equation}
T
=
\frac{1}{4}
T^\rho{}_{\mu\nu}T_\rho{}^{\mu\nu}
+
\frac{1}{2}
T^\rho{}_{\mu\nu}T^{\nu\mu}{}_\rho
-
T^\rho{}_{\mu\rho}T^{\nu\mu}{}_\nu .
\label{eq:torsion_scalar_explicit}
\end{equation}

The action of the teleparallel equivalent of General Relativity (TEGR) is
\begin{equation}
S_{\rm TG}
=
-\frac{1}{2\kappa^2}
\int d^4x\,h\,T,
\label{eq:TEGR_action}
\end{equation} where $\kappa^2=8\pi G$. Also, 
with the conventions adopted here, the torsion scalar and the Ricci scalar
$\mathring{R}$ of the Levi-Civita connection satisfy
\begin{equation}
\mathring{R}
=
-T+
\frac{2}{h}
\partial_\mu
\left(
hT^\nu{}_{\nu}{}^\mu
\right).
\label{eq:R_T_relation}
\end{equation}
Hence, the TEGR and Einstein--Hilbert actions differ only by a boundary term
and yield equivalent gravitational field equations. This equivalence,
however, does not generally extend to modified theories of gravity. Once the
gravitational or matter sectors are modified through nontrivial functions or
couplings involving torsional quantities, the boundary term relating $T$ and
$\mathring{R}$ can no longer, in general, be discarded without affecting the
dynamics. Torsion-based modifications, therefore, define gravitational theories
that are generically distinct from their curvature-based analogs \cite{Cai:2015emx}.

The symmetry structure discussed above is central to the non-universal extension developed below. In the present construction, the departure from universal coupling arises from a field-dependent coupling of the scalar sector to the translational gauge potential, while the pure gravitational sector remains that of TEGR. This coupling leads to a modified gravity-matter system whose consistent implementation must preserve diffeomorphism and local Lorentz covariance, together with local translational gauge invariance.

\section{Non-universal gravitational couplings in teleparallel gravity}
\label{sec:nonuniversal_teleparallel}

The geometrical description of General Relativity relies on the universality of free fall. Since all particles follow the same spacetime geodesics, gravity can be encoded in a single metric. If different particles coupled to gravity with different strengths, their trajectories could not be described as geodesics of the same geometry.

The situation is different in teleparallel gravity. Since gravitation is described as a gauge force, the gravitational field may remain universal even when its coupling to matter is not. This possibility was discussed by Aldrovandi and Pereira by considering a spinless particle with inertial mass $m_i$ and gravitational mass $m_g$ \cite{AldrovandiPereiraBook,Aldrovandi:2003pa}.

The particle action is
\begin{equation}
S_{\rm p}
=
-m_i c
\int d\sigma
-
m_g c
\int
B^{A}{}_{\mu}u_{A}\,dx^{\mu},
\label{eq:nonuniversal_particle_action}
\end{equation}
where
\begin{equation}
d\sigma
=
\left(
\eta_{AB}dx^{A}dx^{B}
\right)^{1/2},
\qquad
u^{A}
=
\frac{dx^{A}}{d\sigma}.
\label{eq:tangent_interval_velocity}
\end{equation}
The first term describes the free particle, whereas the second represents its coupling to the translational gauge potential.

Variation of Eq.~\eqref{eq:nonuniversal_particle_action} with respect to the particle trajectory gives the force equation
\begin{equation}
\left(
\partial_{\mu}x^{A}
+
\frac{m_g}{m_i}
B^{A}{}_{\mu}
\right)
\frac{du_{A}}{ds}
=
\frac{m_g}{m_i}
F^{A}{}_{\mu\nu}
u_{A}u^{\nu}.
\label{eq:nonuniversal_force_equation}
\end{equation}
The dynamics depend on the dimensionless ratio
\begin{equation}
q_g
\equiv
\frac{m_g}{m_i},
\label{eq:gravitational_charge}
\end{equation}
which plays the role of the gravitational charge of the particle. The gauge potential $B^{A}{}_{\mu}$, however, is independent of $q_g$. Particles with different values of $m_g/m_i$ therefore experience different gravitational forces while interacting with the same gravitational field.

One may formally introduce the particle-dependent tetrad
\begin{equation}
h^{A}{}_{\mu}(q_g)
=
D_{\mu}x^{A}
+
q_g B^{A}{}_{\mu}.
\label{eq:particle_dependent_tetrad}
\end{equation}
A metric constructed from this tetrad would depend explicitly on the properties of the test particle,
\begin{equation}
g_{\mu\nu}(q_g)
=
\eta_{AB}
h^{A}{}_{\mu}(q_g)
h^{B}{}_{\nu}(q_g).
\label{eq:particle_dependent_metric}
\end{equation}
Hence, when $q_g$ is not universal, no single metric can geometrize the motion of all particles. The teleparallel force equation remains well defined because the universal field $B^{A}{}_{\mu}$ is unchanged and the non-universality appears only through the matter coupling.

This argument suggests a direct field-theoretical extension. Field-dependent
gravitational couplings arise naturally in scalar--tensor and dilaton
scenarios, as illustrated by the Damour--Polyakov mechanism
\cite{Damour:1994zq,Damour:2002mi}. Motivated by this possibility, we promote
the effective gravitational charge of the scalar sector to a field-dependent
quantity,
\begin{equation}
q_g
\longrightarrow
q_{\phi}(\phi)
=
1+\alpha(\phi),
\label{eq:field_dependent_charge}
\end{equation}
where $\alpha(\phi)$ is a dimensionless function parametrizing the departure
from universality.

In the construction considered below, the gravitational sector remains
described by the universal tetrad $h^{A}{}_{\mu}$, while only the scalar field
couples non-universally to the translational gauge potential. The limit
$\alpha(\phi)=0$ restores the universal teleparallel coupling.

\section{Scalar field with a non-universal gravitational coupling}
\label{sec:scalar_nonuniversal}

We consider a canonical scalar field with a non-universal coupling to the
translational gravitational potential. To preserve local translational gauge
invariance, we introduce four St\"uckelberg fields $\pi^{A}(x)$ transforming
as
\begin{equation}
\delta_{\epsilon}\pi^{A}=\epsilon^{A},
\qquad
\delta_{\epsilon}B^{A}{}_{\mu}
=
-D_{\mu}\epsilon^{A}.
\label{eq:gauge_transformations}
\end{equation}
The combination
\begin{equation}
\mathcal{B}^{A}{}_{\mu}
\equiv
B^{A}{}_{\mu}
+
D_{\mu}\pi^{A}
\label{eq:Bcal_definition}
\end{equation}
is therefore invariant under local translations,
\begin{equation}
\delta_{\epsilon}\mathcal{B}^{A}{}_{\mu}=0.
\end{equation}
Since the gravitational tetrad
\begin{equation}
h^{A}{}_{\mu}
=
e^{A}{}_{\mu}
+
B^{A}{}_{\mu},
\qquad
e^{A}{}_{\mu}=D_{\mu}x^{A},
\end{equation}
is also translationally invariant, the scalar sector can consistently be
coupled to the effective tetrad
\begin{equation}
h_{\phi}^{A}{}_{\mu}
=
h^{A}{}_{\mu}
+
\alpha(\phi)\mathcal{B}^{A}{}_{\mu},
\label{eq:effective_tetrad_scalar}
\end{equation}
where $\alpha(\phi)$ is a dimensionless function. Assuming that $\phi$ is
invariant under local translations, one has
\begin{equation}
\delta_{\epsilon}h_{\phi}^{A}{}_{\mu}=0.
\end{equation}
Thus, a field-dependent non-universal coupling can be introduced without
breaking the local translational gauge symmetry.

For convenience, we define
\begin{equation}
b^{\mu}{}_{\nu}
\equiv
h_{A}{}^{\mu}\mathcal{B}^{A}{}_{\nu},
\qquad
M^{\mu}{}_{\nu}
\equiv
\delta^{\mu}{}_{\nu}
+
\alpha(\phi)b^{\mu}{}_{\nu},
\label{eq:M_b_definitions}
\end{equation}
such that
\begin{equation}
h_{\phi}^{A}{}_{\mu}
=
h^{A}{}_{\nu}M^{\nu}{}_{\mu}.
\label{eq:effective_tetrad_M}
\end{equation}
For an invertible $M^{\mu}{}_{\nu}$, the inverse effective tetrad is
\begin{equation}
h_{\phi A}{}^{\mu}
=
(M^{-1})^{\mu}{}_{\nu}h_{A}{}^{\nu},
\label{eq:inverse_effective_tetrad}
\end{equation}
and the corresponding effective metric and its inverse are
\begin{align}
g_{\mu\nu}^{(\phi)}
&=
g_{\rho\sigma}
M^{\rho}{}_{\mu}
M^{\sigma}{}_{\nu},
\label{eq:effective_metric_M}
\\
g_{(\phi)}^{\mu\nu}
&=
(M^{-1})^{\mu}{}_{\rho}
(M^{-1})^{\nu}{}_{\sigma}
g^{\rho\sigma}.
\label{eq:inverse_effective_metric_M}
\end{align}
The effective volume element satisfies
\begin{equation}
h_{\phi}
=
h\,\det M,
\label{eq:effective_determinant}
\end{equation}
where $h=\det(h^{A}{}_{\mu})=\sqrt{-g}$. We restrict the theory to the
nondegenerate branch $\det M>0$.

The scalar-field action is
\begin{equation}
S_{\phi}
=
\int d^{4}x\,h_{\phi}
\left[
-\frac12
g_{(\phi)}^{\mu\nu}
\partial_{\mu}\phi\partial_{\nu}\phi
-
V(\phi)
\right],
\label{eq:scalar_action_eff}
\end{equation}
or, equivalently,
{\smaller
\begin{equation}
\begin{aligned}
S_{\phi}
=
\int d^{4}x\,h\,\det M
\bigg[
&-\frac12
g^{\rho\sigma}
(M^{-1})^{\mu}{}_{\rho}
(M^{-1})^{\nu}{}_{\sigma}
\partial_{\mu}\phi\partial_{\nu}\phi
\\
&-V(\phi)
\bigg].
\end{aligned}
\label{eq:exact_scalar_action}
\end{equation}
No perturbative expansion in $\alpha(\phi)$ is assumed.

The complete action is therefore
\bea
S=S_{\rm TG}[h]+S_{\phi}[h_{\phi},\phi],
    \label{eq:complete_action}
\eea
where $S_{\rm TG}[h]$ is the gravitational action given in
Eq.~\eqref{eq:TEGR_action}, and $S_{\phi}[h_{\phi},\phi]$ is the scalar-field action with non-universal gravitational coupling defined in
Eq.~\eqref{eq:exact_scalar_action}. The gravitational sector is governed by the universal tetrad
$h^{A}{}_{\mu}$, whereas the scalar field probes the effective tetrad
$h_{\phi}^{A}{}_{\mu}$ induced by its field-dependent coupling to the
gauge-invariant translational sector $\mathcal{B}^{A}{}_{\mu}$.

In the unitary gauge
\begin{equation}
\pi^{A}=0,
\label{eq:unitary_gauge}
\end{equation}
the gauge-invariant combination reduces to
\begin{equation}
\mathcal{B}^{A}{}_{\mu}=B^{A}{}_{\mu},
\end{equation}
and the effective tetrad becomes
\begin{equation}
h_{\phi}^{A}{}_{\mu}
=
h^{A}{}_{\mu}
+
\alpha(\phi)B^{A}{}_{\mu}.
\label{eq:effective_tetrad_unitary}
\end{equation}
Accordingly,
\begin{equation}
b^{\mu}{}_{\nu}
=
h_{A}{}^{\mu}B^{A}{}_{\nu},
\qquad
M^{\mu}{}_{\nu}
=
\delta^{\mu}{}_{\nu}
+
\alpha(\phi)b^{\mu}{}_{\nu}.
\label{eq:M_unitary}
\end{equation}
Equation~\eqref{eq:effective_tetrad_unitary} thus represents the
unitary-gauge form of the translationally invariant theory. The universal
limit is recovered for $\alpha(\phi)=0$.

\section{Field equations}
\label{sec:field_equations}

The field equations follow from independent variations of the action
\eqref{eq:complete_action} with respect to $B^{A}{}_{\mu}$, $\pi^{A}$, and
$\phi$. The unitary gauge is imposed only after the equations of motion have
been derived.

\subsection{Gravitational field equations}
\label{subsec:gravitational_equations}

Under a variation of the translational gauge potential, keeping
$\pi^{A}$ and $\phi$ fixed,
\begin{equation}
\delta_{B}h^{A}{}_{\mu}
=
\delta B^{A}{}_{\mu},
\qquad
\delta_{B}\mathcal{B}^{A}{}_{\mu}
=
\delta B^{A}{}_{\mu}.
\end{equation}
It follows directly from the definition of the scalar-sector tetrad that
\begin{equation}
\delta_{B}h_{\phi}^{A}{}_{\mu}
=
\left[1+\alpha(\phi)\right]
\delta B^{A}{}_{\mu}.
\label{eq:delta_heff_B}
\end{equation}

The scalar energy-momentum tensor defined with respect to the effective
metric is
\bea
&& T_{\mu\nu}^{(\phi)}\equiv 
-\frac{2}{h_{\phi}}
\frac{\delta S_{\phi}}{\delta g_{(\phi)}^{\mu\nu}} \nonumber\\
&& 
=
\partial_{\mu}\phi\partial_{\nu}\phi
-
g_{\mu\nu}^{(\phi)}
\left[
\frac12
g_{(\phi)}^{\rho\sigma}
\partial_{\rho}\phi\partial_{\sigma}\phi
+
V(\phi)
\right].
\label{eq:Tphi}
\eea
Introducing
\begin{equation}
\Theta_{A}^{(\phi)\mu}
\equiv 
h_{\phi A}{}^{\nu}
T^{(\phi)\mu}{}_{\nu},
\label{eq:Theta_phi}
\end{equation}
the scalar contribution to the variation becomes
\begin{equation}
\delta_{B}S_{\phi}
=
\int d^{4}x\,h
\left[1+\alpha(\phi)\right]
\det M\,
\Theta_{A}^{(\phi)\mu}
\delta B^{A}{}_{\mu}.
\label{eq:delta_scalar_B}
\end{equation}

The TEGR contribution can be written as
\begin{equation}
\delta_{B}S_{\rm TG}
=
-\frac{1}{\kappa^{2}}
\int d^{4}x\,h\,
\mathcal{E}_{A}{}^{\mu}
\delta B^{A}{}_{\mu},
\end{equation}
where
\begin{equation}
\mathcal{E}_{A}{}^{\mu}
\equiv
-\frac{\kappa^{2}}{h}
\frac{\delta S_{\rm TG}}{\delta B^{A}{}_{\mu}}
=
h_{A}{}^{\lambda}\mathring{G}_{\lambda}{}^{\mu},
\label{eq:TEGR_EL}
\end{equation}
and $\mathring{G}_{\mu\nu}$ is the Einstein tensor associated with the
universal metric $g_{\mu\nu}$.

The gravitational field equations are therefore
\begin{equation}
\mathcal{E}_{A}{}^{\mu}
=
\kappa^{2}
\left[1+\alpha(\phi)\right]
\det M\,
\Theta_{A}^{(\phi)\mu}.
\label{eq:gravitational_field_equation_tetrad}
\end{equation}
Using
\begin{equation}
h^{A}{}_{\lambda}
h_{\phi A}{}^{\nu}
=
(M^{-1})^{\nu}{}_{\lambda},
\end{equation}
the equivalent spacetime form is
\begin{equation}
\mathring{G}_{\lambda}{}^{\mu}
=
\kappa^{2}
\left[1+\alpha(\phi)\right]
\det M\,
(M^{-1})^{\nu}{}_{\lambda}
T^{(\phi)\mu}{}_{\nu}.
\label{eq:gravitational_field_equation}
\end{equation}
The effective source for the universal gravitational tetrad is thus
\begin{equation}
T_{\lambda}^{\rm eff\,\mu}
=
\left[1+\alpha(\phi)\right]
\det M\,
(M^{-1})^{\nu}{}_{\lambda}
T^{(\phi)\mu}{}_{\nu}.
\label{eq:effective_energy_momentum}
\end{equation}

Since the Einstein tensor satisfies the contracted Bianchi identity, the
gravitational field equations imply
\begin{equation}
\mathring{\nabla}_{\mu}
T_{\rm eff}^{\mu}{}_{\nu}
=
0,
\label{eq:effective_conservation}
\end{equation}
where $\mathring{\nabla}_{\mu}$ is the Levi--Civita covariant derivative
associated with the universal metric $g_{\mu\nu}$. Thus, the effective
energy--momentum tensor is covariantly conserved with respect to the
universal geometry.

The symmetry of the Einstein tensor further requires the antisymmetric part
of the effective source in Eq.~\eqref{eq:gravitational_field_equation} to
vanish. This yields the constraint
\begin{equation}
g_{\rho[\mu}
(M^{-1})^{\nu}{}_{\lambda]}
T^{(\phi)\rho}{}_{\nu}
=
0.
\label{eq:antisymmetric_condition_compact}
\end{equation}
Equation~\eqref{eq:antisymmetric_condition_compact} is the general
antisymmetric constraint implied by the gravitational field equations. No
additional symmetry condition on $M^{\mu}{}_{\nu}$ has been assumed.

\subsection{Scalar-field equation}
\label{subsec:scalar_equation}

The scalar field enters explicitly through its canonical Lagrangian and
implicitly through $\alpha(\phi)$. At fixed $B^{A}{}_{\mu}$ and $\pi^{A}$,
\begin{equation}
\delta_{\phi}h_{\phi}^{A}{}_{\mu}
=
\alpha_{,\phi}
\mathcal{B}^{A}{}_{\mu}\delta\phi .
\label{eq:delta_heff_phi}
\end{equation}

The scalar-field equation is therefore
\begin{equation}
\Box_{(\phi)}\phi
-
V_{,\phi}
+
\alpha_{,\phi}
\mathcal{B}^{A}{}_{\mu}
\Theta_{A}^{(\phi)\mu}
=
0,
\label{eq:scalar_field_equation}
\end{equation}
where
\begin{equation}
\Box_{(\phi)}\phi
\equiv
\frac{1}{h_{\phi}}
\partial_{\mu}
\left(
h_{\phi}
g_{(\phi)}^{\mu\nu}
\partial_{\nu}\phi
\right).
\label{eq:effective_dalembertian}
\end{equation}
The last term in Eq.~\eqref{eq:scalar_field_equation} arises from the
field dependence of the non-universal gravitational coupling.

Using the standard divergence identity for a canonical scalar field,
Eq.~\eqref{eq:scalar_field_equation} implies
\begin{equation}
\nabla^{(\phi)}_{\mu}
T^{(\phi)\mu}{}_{\nu}
=
-
\alpha_{,\phi}
\mathcal{B}^{A}{}_{\mu}
\Theta_{A}^{(\phi)\mu}
\partial_{\nu}\phi .
\label{eq:scalar_nonconservation}
\end{equation}
Thus, for a field-dependent coupling, the scalar energy--momentum tensor is
not separately conserved with respect to the effective geometry.

The same balance law can be expressed in terms of the universal geometry.
We introduce the difference between the corresponding Levi--Civita
connections,
\begin{equation}
C^{\rho}{}_{\mu\nu}
\equiv
\mathring{\Gamma}^{\rho}{}_{\mu\nu}
-
\Gamma^{(\phi)\rho}{}_{\mu\nu},
\label{eq:connection_difference}
\end{equation}
which can be written as
\begin{equation}
C^{\rho}{}_{\mu\nu}
=
-\frac{1}{2}
g_{(\phi)}^{\rho\lambda}
\left(
\mathring{\nabla}_{\mu}g^{(\phi)}_{\lambda\nu}
+
\mathring{\nabla}_{\nu}g^{(\phi)}_{\lambda\mu}
-
\mathring{\nabla}_{\lambda}g^{(\phi)}_{\mu\nu}
\right).
\label{eq:connection_difference_explicit}
\end{equation}
Since $h_{\phi}=h\det M$, its trace satisfies
\begin{equation}
C^{\mu}{}_{\mu\nu}
=
-
\partial_{\nu}\ln\det M .
\label{eq:connection_difference_trace}
\end{equation}
For the scalar energy--momentum tensor, the two covariant derivatives are
related by
\begin{equation}
\mathring{\nabla}_{\mu}
T^{(\phi)\mu}{}_{\nu}
=
\nabla^{(\phi)}_{\mu}
T^{(\phi)\mu}{}_{\nu}
+
C^{\mu}{}_{\mu\rho}
T^{(\phi)\rho}{}_{\nu}
-
C^{\rho}{}_{\mu\nu}
T^{(\phi)\mu}{}_{\rho}.
\label{eq:scalar_divergence_relation}
\end{equation}
Using Eqs.~\eqref{eq:scalar_nonconservation} and
\eqref{eq:connection_difference_trace}, one obtains
\begin{align}
\mathring{\nabla}_{\mu}
T^{(\phi)\mu}{}_{\nu}
={}&
-
\alpha_{,\phi}
\mathcal{B}^{A}{}_{\mu}
\Theta_{A}^{(\phi)\mu}
\partial_{\nu}\phi
\nonumber\\
&-
\left(
\partial_{\rho}\ln\det M
\right)
T^{(\phi)\rho}{}_{\nu}
-
C^{\rho}{}_{\mu\nu}
T^{(\phi)\mu}{}_{\rho}.
\label{eq:scalar_nonconservation_universal}
\end{align}
Equations~\eqref{eq:scalar_nonconservation} and
\eqref{eq:scalar_nonconservation_universal} describe the same
scalar-sector energy-momentum exchange in the effective and universal
geometries, respectively. In general, $T^{(\phi)\mu}{}_{\nu}$ is therefore
not separately conserved with respect to either connection. Its consistency
with the conserved effective source is discussed together with the
St\"uckelberg equations below.

\subsection{St\"uckelberg equations and consistency relations}
\label{subsec:stueckelberg_equations}

The St\"uckelberg fields enter the scalar action through
$\mathcal{B}^{A}{}_{\mu}$. Their variation gives
\begin{equation}
\delta_{\pi}\mathcal{B}^{A}{}_{\mu}
=
D_{\mu}\delta\pi^{A},
\qquad
\delta_{\pi}h_{\phi}^{A}{}_{\mu}
=
\alpha(\phi)D_{\mu}\delta\pi^{A}.
\end{equation}
Variation of the scalar action and integration by parts then yield
\begin{equation}
D_{\mu}
\left[
h\,\alpha(\phi)\det M\,
\Theta_{A}^{(\phi)\mu}
\right]
=
0.
\label{eq:Stueckelberg_equation_M}
\end{equation}
The St\"uckelberg equations provide the constraints associated with local
translational gauge invariance. They are not independent of the remaining
field equations, but are related to them through the corresponding Noether
identity.

The consistency between the scalar and gravitational sectors can be made
explicit by introducing the tensor
\begin{equation}
D^{\rho}{}_{\mu\nu}
\equiv
(M^{-1})^{\rho}{}_{\sigma}
\mathring{\nabla}_{\mu}M^{\sigma}{}_{\nu}
+
C^{\rho}{}_{\mu\nu}.
\label{eq:D_tensor}
\end{equation}
It characterizes the relative deformation between the universal and
effective geometries and vanishes in the universal limit.

Using Eq.~\eqref{eq:effective_energy_momentum}, the universal divergence of
the effective source satisfies the geometric identity
\begin{align}
\mathring{\nabla}_{\mu}
T_{\rm eff}^{\mu}{}_{\nu}
={}&
\left[1+\alpha(\phi)\right]
\det M\,
(M^{-1})^{\rho}{}_{\nu}
\Bigg[
\nabla^{(\phi)}_{\mu}
T^{(\phi)\mu}{}_{\rho}
\nonumber\\
&+
\frac{\alpha_{,\phi}}{1+\alpha}
T^{(\phi)\mu}{}_{\rho}
\partial_{\mu}\phi
-
T^{(\phi)\mu}{}_{\sigma}
D^{\sigma}{}_{\mu\rho}
\Bigg].
\label{eq:conservation_master_identity}
\end{align}
No field equation has been used in deriving
Eq.~\eqref{eq:conservation_master_identity}.

The gravitational field equations and the contracted Bianchi identity imply
Eq.~\eqref{eq:effective_conservation}. Comparing the resulting scalar
balance relation with Eq.~\eqref{eq:scalar_nonconservation} gives
\begin{align}
0={}&
\alpha_{,\phi}
\mathcal{B}^{A}{}_{\mu}
\Theta_{A}^{(\phi)\mu}
\partial_{\nu}\phi
-
\frac{\alpha_{,\phi}}{1+\alpha}
T^{(\phi)\mu}{}_{\nu}
\partial_{\mu}\phi
\nonumber\\
&+
T^{(\phi)\mu}{}_{\rho}
D^{\rho}{}_{\mu\nu}.
\label{eq:conservation_consistency}
\end{align}

Equation~\eqref{eq:conservation_consistency} is the on-shell compatibility
condition between the scalar-field and gravitational equations. Through the
Noether identity associated with local translational invariance, it is
equivalent to the St\"uckelberg equation
\eqref{eq:Stueckelberg_equation_M} once the remaining field equations are
imposed. Thus, it does not constitute an additional independent constraint,
but represents the spacetime form of the same St\"uckelberg condition.

Thus, the non-conservation of the scalar energy-momentum tensor in the effective and universal geometries, and the conservation of the effective source in the universal geometry, form a mutually consistent system related through the St\"uckelberg/Noether sector. In the universal limit, the
standard conservation law
$\mathring{\nabla}_{\mu}T^{(\phi)\mu}{}_{\nu}=0$ is recovered.

The St\"uckelberg equations must nevertheless be retained when the unitary
gauge, $\pi^{A}=0$, is imposed in the non-universal theory.

\section{Homogeneous and isotropic cosmology}
\label{sec:cosmology}

As a first application, we consider a spatially flat
Friedmann--Lema\^{i}tre--Robertson--Walker spacetime,
\begin{equation}
ds^{2}
=
-dt^{2}
+
a^{2}(t)\delta_{ij}dx^{i}dx^{j},
\label{eq:FLRW_metric}
\end{equation}
described by the diagonal tetrad
\begin{equation}
h^{A}{}_{\mu}
=
{\rm diag}(1,a,a,a).
\label{eq:FLRW_tetrad}
\end{equation} This is a proper tetrad, and the associated spin connection is $\omega^{A}_{~B \mu}=0$ \cite{Krssak:2015oua}. 
We work in unitary gauge, $\pi^{A}=0$, after deriving the complete
gauge-invariant field equations. In Cartesian coordinates, the inertial
tetrad is chosen as
\begin{equation}
e^{A}{}_{\mu}
=
{\rm diag}(1,1,1,1),
\end{equation}
so that
\begin{equation}
B^{A}{}_{\mu}
=
{\rm diag}(0,a-1,a-1,a-1).
\label{eq:FLRW_B}
\end{equation}

For this background,
\begin{equation}
b^{\mu}{}_{\nu}
=
h_{A}{}^{\mu}B^{A}{}_{\nu}
=
{\rm diag}(0,\beta,\beta,\beta),
\label{eq:FLRW_b}
\end{equation} where $\beta(a)
\equiv
1-\frac{1}{a}$. 

We introduce the deformation factor
\begin{equation}
m(a,\phi)
\equiv
1+\alpha(\phi)\beta(a)
=
1+\alpha(\phi)
\left(
1-\frac{1}{a}
\right).
\label{eq:m_definition}
\end{equation}
The matrix $M^{\mu}{}_{\nu}$ and its inverse are then
\bea
&& M^{\mu}{}_{\nu}
=
{\rm diag}(1,m,m,m),
\nonumber\\
&& 
(M^{-1})^{\mu}{}_{\nu}
=
{\rm diag}(1,m^{-1},m^{-1},m^{-1}),
\label{eq:FLRW_M}
\eea
with $\det M=m^{3}$. For the proper FLRW tetrad, the translational gauge potential has
$B^{0}{}_{0}=0$, and therefore $M^{0}{}_{0}=1$. Hence, the non-universal
deformation affects only the spatial components of the scalar-sector tetrad.
We further require $m>0$ to ensure a nondegenerate effective tetrad.

The scalar-sector tetrad becomes
\begin{equation}
h_{\phi}^{A}{}_{\mu}
=
{\rm diag}(1,a_{\phi},a_{\phi},a_{\phi}),
\end{equation}
where
\begin{equation}
a_{\phi}
=
am
=
a+\alpha(\phi)(a-1).
\label{eq:effective_scale_factor}
\end{equation}
The scalar field therefore propagates on the effective geometry
\begin{equation}
ds_{\phi}^{2}
=
-dt^{2}
+
a_{\phi}^{2}(t)\delta_{ij}dx^{i}dx^{j}.
\label{eq:effective_FLRW_metric}
\end{equation}
The corresponding Hubble parameter is
\begin{equation}
H_{\phi}
\equiv
\frac{\dot a_{\phi}}{a_{\phi}}
=
\frac{1+\alpha}{m}H
+
\frac{\alpha_{,\phi}\beta}{m}\dot\phi,
\label{eq:Heff}
\end{equation}
where $H=\dot a/a$.

For a homogeneous scalar field, $\phi=\phi(t)$,
\begin{align}
\rho_{\phi}
&=
\frac{1}{2}\dot\phi^{2}+V(\phi),
\label{eq:rho_phi}
\\
p_{\phi}
&=
\frac{1}{2}\dot\phi^{2}-V(\phi).
\label{eq:p_phi}
\end{align}

For the diagonal FLRW ansatz, both $M^{\mu}{}_{\nu}$ and
$T^{(\phi)\mu}{}_{\nu}$ are diagonal. Hence, the antisymmetric part of the
effective gravitational source vanishes,
\begin{equation}
T^{\rm eff}_{[\mu\nu]}=0,
\label{eq:FLRW_antisymmetric}
\end{equation}
and the antisymmetric gravitational equations are automatically satisfied
at the homogeneous and isotropic level.

The temporal component of the gravitational field equations gives
\begin{equation}
3H^{2}
=
\kappa^{2}
[1+\alpha(\phi)]
m^{3}\rho_{\phi},
\label{eq:first_Friedmann_NU}
\end{equation}
whereas the spatial components yield
\begin{equation}
2\dot H+3H^{2}
=
-\kappa^{2}
[1+\alpha(\phi)]
m^{2}p_{\phi}.
\label{eq:second_Friedmann_NU}
\end{equation}
Equivalently,
\begin{equation}
\dot H
=
-\frac{\kappa^{2}}{2}
[1+\alpha(\phi)]
m^{2}
\left(
m\rho_{\phi}+p_{\phi}
\right).
\label{eq:Hdot_NU}
\end{equation}

Defining
\begin{align}
\rho_{\rm eff}
&=
[1+\alpha(\phi)]m^{3}\rho_{\phi},
\label{eq:rho_eff_cosmo}
\\
p_{\rm eff}
&=
[1+\alpha(\phi)]m^{2}p_{\phi},
\label{eq:p_eff_cosmo}
\end{align}
the gravitational equations take the standard form
\begin{equation}
3H^{2}
=
\kappa^{2}\rho_{\rm eff},
\qquad
2\dot H+3H^{2}
=
-\kappa^{2}p_{\rm eff}.
\end{equation}
The corresponding effective equation-of-state parameter is
\begin{equation}
w_{\rm eff}
=
\frac{p_{\rm eff}}{\rho_{\rm eff}}
=
\frac{1}{m}
\frac{p_{\phi}}{\rho_{\phi}}.
\label{eq:weff_general}
\end{equation}
The first slow-roll parameter can be written exactly as
\begin{equation}
\epsilon
\equiv
-\frac{\dot H}{H^{2}}
=
\frac{3}{2}
\frac{m\rho_{\phi}+p_{\phi}}
     {m\rho_{\phi}}.
\label{eq:epsilon_general}
\end{equation}
Using Eqs.~\eqref{eq:rho_phi} and \eqref{eq:p_phi}, this becomes
\begin{equation}
\epsilon
=
\frac{3}{2}
\frac{
(m+1)\dot\phi^{2}/2
+
(m-1)V
}{
m\left(\dot\phi^{2}/2+V\right)
}.
\label{eq:epsilon_general_scalar}
\end{equation}

The homogeneous scalar equation is
\begin{equation}
\ddot\phi
+
3H_{\phi}\dot\phi
+
V_{,\phi}
+
3\frac{\alpha_{,\phi}\beta}{m}
\left(
V-\frac{1}{2}\dot\phi^{2}
\right)
=
0.
\label{eq:scalar_FLRW}
\end{equation}
Equivalently,
\begin{equation}
\dot\rho_{\phi}
+
3H_{\phi}
(\rho_{\phi}+p_{\phi})
-
3\frac{\alpha_{,\phi}\beta}{m}
p_{\phi}\dot\phi
=
0.
\label{eq:scalar_continuity_FLRW}
\end{equation}

The temporal component of the St\"uckelberg equation gives
\begin{equation}
\frac{d}{dt}
\left[
\alpha(\phi)a_{\phi}^{3}\rho_{\phi}
\right]
=
0,
\label{eq:Stueckelberg_FLRW}
\end{equation}
and therefore
\begin{equation}
\alpha(\phi)a_{\phi}^{3}\rho_{\phi}
=
{\cal C},
\label{eq:Stueckelberg_first_integral}
\end{equation}
where ${\cal C}$ is an integration constant. The spatial components vanish
identically due to homogeneity and isotropy. Moreover,
\begin{equation}
\dot\rho_{\rm eff}
+
3H(\rho_{\rm eff}+p_{\rm eff})
=
\frac{1}{a^{3}}
\frac{d}{dt}
\left[
\alpha(\phi)a_{\phi}^{3}\rho_{\phi}
\right]=0.
\label{eq:FLRW_consistency_relation}
\end{equation}
Thus, the St\"uckelberg relation provides the background consistency
condition connecting the scalar and gravitational equations.

\subsection{Kination-dominated regime}
\label{subsec:kination}
We first consider the kinetic-energy-dominated regime,
\begin{equation}
\frac{1}{2}\dot{\phi}^{2}\gg V(\phi),
\label{eq:kin_regime}
\end{equation}
which may describe an early stage preceding potential-dominated inflation. At leading order, $\rho_{\phi}\simeq p_{\phi}\simeq\dot{\phi}^{2}/2$, and the background equations reduce to
\bea
&& 3H^{2} \simeq \frac{\kappa^{2}}{2}(1+\alpha)m^{3}\dot{\phi}^{2},
\label{eq:kin_Friedmann}\\
&& \dot H \simeq -\frac{\kappa^{2}}{4}(1+\alpha)m^{2}(m+1)\dot{\phi}^{2},
\label{eq:kin_Hdot}\\
&& \ddot{\phi}+3H_{\phi}\dot{\phi} \simeq \frac{3}{2}\frac{\alpha_{,\phi}\beta}{m}\dot{\phi}^{2},
\label{eq:kin_scalar}
\eea
while the St\"uckelberg constraint becomes
\begin{equation}
\alpha a_{\phi}^{3}\dot{\phi}^{2}=C_{K}.
\label{eq:kin_Stueckelberg}
\end{equation}
Since Eq.~\eqref{eq:kin_scalar} follows from the other three through the background consistency identity, we integrate Eqs.~\eqref{eq:kin_Friedmann}, \eqref{eq:kin_Hdot}, and \eqref{eq:kin_Stueckelberg}. The ratio of Eqs.~\eqref{eq:kin_Hdot} and \eqref{eq:kin_Friedmann} gives
\begin{equation}
\frac{d\ln H}{d\ln a}\simeq-\frac{3}{2}\left(1+\frac{1}{m}\right),
\label{eq:kin_dlnH}
\end{equation}
while eliminating $\dot{\phi}^{2}$ between Eqs.~\eqref{eq:kin_Friedmann} and \eqref{eq:kin_Stueckelberg} yields
\begin{equation}
\frac{\alpha}{1+\alpha}\,a^{3}H^{2}=C,
\qquad
C\equiv\frac{\kappa^{2}}{6}C_{K}.
\label{eq:kin_first_integral}
\end{equation}
Differentiating Eq.~\eqref{eq:kin_first_integral} and using Eq.~\eqref{eq:kin_dlnH} gives
\begin{equation}
\frac{d\alpha}{d\ln a}\simeq\frac{3\alpha(1+\alpha)}{m},
\label{eq:kin_dalpha}
\end{equation}
so that the evolution of the non-universal coupling is fixed by the background equations. Equation~\eqref{eq:kin_dalpha} integrates to
\begin{equation}
a\,\alpha^{-1/3}=C_{1}-\frac{1}{3}\int^{\alpha}\frac{x^{-1/3}}{1+x}\,dx ,
\label{eq:kin_alpha_implicit}
\end{equation}
and, for $C_{1}\neq0$, its small-$\alpha$ expansion gives
\begin{equation}
\alpha(a)\simeq Aa^{3},
\qquad
m(a)\simeq1-Aa^{2}+\mathcal{O}(a^{3}),
\label{eq:kin_alpha_m}
\end{equation}
where $A\equiv C_{1}^{-3}$. The universal geometry is thus recovered as $a\to0$, and strong coupling is excluded, since $m>0$ requires $-1<\alpha<a/(1-a)$ for $a<1$. Integrating Eq.~\eqref{eq:kin_dlnH} then gives
\begin{equation}
H(a)\simeq H_{*}\,a^{-3}\left[1-\frac{3}{4}Aa^{2}+\mathcal{O}(a^{3})\right],
\label{eq:kin_H}
\end{equation}
with $H_{*}^{2}=C/A$ by Eq.~\eqref{eq:kin_first_integral}. The leading term is the stiff-fluid solution, and $w_{\rm eff}=1/m\simeq1+Aa^{2}$. Similarly, Eq.~\eqref{eq:kin_Friedmann} gives
\begin{equation}
\phi(a)\simeq\phi_{0}\pm\frac{\sqrt{6}}{\kappa}\left[\ln a+\frac{3}{4}Aa^{2}+\mathcal{O}(a^{3})\right],
\label{eq:kin_phi}
\end{equation}
where $\phi_{0}$ is a constant. Using Eqs.~(85) and \eqref{eq:kin_dalpha}, one verifies that Eq.~\eqref{eq:kin_phi} satisfies Eq.~\eqref{eq:kin_scalar}. Combining Eqs.~\eqref{eq:kin_alpha_m} and \eqref{eq:kin_phi}, the coupling function must behave asymptotically as
\begin{equation}
\alpha(\phi)\simeq A\exp\left[\pm\sqrt{\frac{3}{2}}\,\kappa(\phi-\phi_{0})\right].
\label{eq:kin_alpha_phi}
\end{equation}
The approximation~\eqref{eq:kin_regime} holds provided $V/\dot{\phi}^{2}$ and $V_{,\phi}/(H\dot{\phi})$ vanish as $a\to0$. This solution thus describes a pre-inflationary kination phase that approaches the universal stiff solution as $a\to0$.

\subsection{Potential-dominated regime}
\label{sec:potential_domination}

We investigate whether the scalar field can drive accelerated expansion
when its potential energy dominates
\begin{equation}
r_{\phi}\equiv\frac{\dot\phi^{2}}{2V(\phi)}\ll1,
\label{eq:pd_ratio}
\end{equation}
without imposing the second conventional slow-roll condition.
Using $w_{\phi}=(r_{\phi}-1)/(r_{\phi}+1)$ in Eq.~\eqref{eq:epsilon_general}, we obtain
\begin{equation}
\epsilon=\frac32\left[1-\frac1m+
\frac{2r_{\phi}}{m(1+r_{\phi})}\right].
\label{eq:pd_epsilon}
\end{equation}
For $m>0$ and $V>0$, accelerated expansion ($\epsilon<1$)
requires
\begin{equation}
r_{\phi}<\frac{3-m}{3+m},\qquad 0<m<3.
\label{eq:pd_acceleration}
\end{equation}
Since $m=a_{\phi}/a$, the limit $m\to1$ recovers the standard
condition $r_{\phi}<1/2$, whereas $m\to3$ marks the boundary
beyond which acceleration is impossible. Thus, potential domination
alone need not imply acceleration for arbitrary spatial deformation.

The St\"uckelberg first integral \eqref{eq:Stueckelberg_first_integral} and the Friedmann equation \eqref{eq:first_Friedmann_NU}
give the exact identity
\begin{equation}
\frac{\alpha}{1+\alpha}\,a^{3}H^{2}
=\frac{\kappa^{2}\mathcal C}{3}.
\label{eq:pd_first_integral}
\end{equation}
Taking its logarithmic derivative and using Eq.~\eqref{eq:pd_epsilon} yields
\begin{equation}
\frac{d\alpha}{dN}
=3\frac{r_{\phi}-1}{r_{\phi}+1}
\frac{\alpha(1+\alpha)}{m},
\qquad N\equiv\ln a.
\label{eq:pd_alpha_evolution}
\end{equation}
For $r_{\phi}<1$, $m>0$, and $1+\alpha>0$, the magnitude
$|\alpha|$ decreases during expansion. Writing $d\alpha/dN=\alpha_{,\phi}\dot\phi/H$ in Eq. \eqref{eq:pd_alpha_evolution} and using Eq.~\eqref{eq:first_Friedmann_NU} to express $\dot\phi^2/H^2$ in terms of $r_\phi$ as defined in Eq. \eqref{eq:pd_ratio}, we obtain
\begin{equation}
\frac{r_{\phi}(1+r_{\phi})}{(1-r_{\phi})^{2}}
=\frac{3\kappa^{2}}{2}
\frac{(1+\alpha)^{3}m}
{\bigl(\partial_{\phi}\ln|\alpha|\bigr)^{2}}.
\label{eq:pd_exact_ratio}
\end{equation}
This exact relation holds on the nontrivial rolling branch
$\alpha\ne0$, $\alpha_{,\phi}\ne0$. On $0\le r_{\phi}<1$,
Eqs.~\eqref{eq:pd_acceleration} and \eqref{eq:pd_exact_ratio}
imply the acceleration criterion
\begin{equation}
\bigl(\partial_{\phi}\ln|\alpha|\bigr)^{2}
>\frac{\kappa^{2}(1+\alpha)^{3}m^{3}}{3-m},
\qquad 0<m<3.
\label{eq:pd_acceleration_slope}
\end{equation}
Potential domination is realized on this branch when
\begin{equation}
\bigl(\partial_{\phi}\ln|\alpha|\bigr)^{2}
\gg\frac{3\kappa^{2}}2(1+\alpha)^{3}m,
\label{eq:pd_domination_slope}
\end{equation}
for which
\begin{equation}
r_{\phi}\simeq\frac{3\kappa^{2}}2
\frac{(1+\alpha)^{3}m}
{\bigl(\partial_{\phi}\ln|\alpha|\bigr)^{2}}\ll1.
\label{eq:pd_ratio_approx}
\end{equation}
Thus, potential domination is controlled by the logarithmic slope
of the coupling, not by $m$ alone. The compatible potential along
the background is constrained by Eq.~\eqref{eq:Stueckelberg_first_integral}.
Together with the gravitational equations, the scalar-field equation
\eqref{eq:scalar_FLRW} is consistent through the identity \eqref{eq:FLRW_consistency_relation} for $\dot\phi\ne0$.

Equation~\eqref{eq:pd_first_integral} further implies
$\alpha\propto a^{-3}H^{-2}$ when $|\alpha|\ll1$.
If potential domination leads to an asymptotically constant $H$,
then $\alpha\propto a^{-3}$. Two coupling functions illustrate
how the resulting expansion depends on the scalar dynamics.

For an exponential coupling,
\begin{equation}
\alpha(\phi)=A e^{\lambda\kappa\phi},
\label{eq:pd_exponential}
\end{equation}
Eq.~\eqref{eq:pd_exact_ratio} implies that, as $\alpha\to0$,
$r_{\phi}$ approaches a constant satisfying
\begin{equation}
\frac{r_{\phi}(1+r_{\phi})}{(1-r_{\phi})^2}
\longrightarrow\frac{3}{2\lambda^2}.
\label{eq:pd_exponential_ratio}
\end{equation}
On the branch $0\le r_{\phi}<1$, accelerated expansion occurs
for $\lambda^2>1/2$, whereas potential domination requires
$\lambda^2\gg3/2$. In the latter regime,
$r_{\phi}\simeq3/(2\lambda^2)$ and
$\epsilon\simeq9/(2\lambda^2)$, corresponding to asymptotic
power-law inflation. Accordingly, the coupling scales as
$\alpha\propto a^{-3+2\epsilon}$, approaching $a^{-3}$
in the quasi-de Sitter limit.

Alternatively, consider a power-law coupling,
\begin{equation}
\alpha(\phi)=A(\phi-\phi_I)^n,\qquad n>0.
\label{eq:pd_power_law}
\end{equation}
Along a compatible expanding solution approaching $\phi_I$,
where $\alpha\to0$ and $m\to1$, Eq.~\eqref{eq:pd_ratio_approx}
yields
\begin{equation}
r_{\phi}\simeq\frac{3\kappa^{2}}{2n^{2}}
(\phi-\phi_I)^{2}\propto a^{-6/n}\longrightarrow0.
\label{eq:pd_power_law_ratio}
\end{equation}
Consequently, $\epsilon\to0$: both potential domination and
accelerated expansion become asymptotically self-consistent.
For $n=2$, the coupling
$\alpha\simeq\kappa^{2}(\phi-\phi_I)^2/8$
is compatible with an asymptotically constant potential and gives
$r_{\phi}\simeq3\alpha\to0$. This explicitly illustrates
potential-driven acceleration without imposing
$|\ddot\phi|\ll|3H_{\phi}\dot\phi|$.

\subsection{Universal matter and a late-time cosmological solution}
\label{sec:late_time}

We now introduce ordinary matter universally coupled to the gravitational
 tetrad $h^A{}_{\mu}$, while the scalar field remains coupled to
$h^A_{\phi\mu}$. The action is extended to
\begin{equation}
S=S_{\rm TG}[h]+S_\phi[h_\phi,\phi]+S_m[h,\Psi_m],
\label{eq:late_action}
\end{equation}
where $\Psi_m$ denotes the matter fields. For a perfect fluid with
constant equation-of-state parameter $w$, universal minimal coupling gives
\begin{equation}
p_m=w\rho_m,\qquad \rho_m=\rho_{m0}a^{-3(1+w)}.
\label{eq:late_matter}
\end{equation}
Here $\rho_{m0}$ is an integration constant in the scale-factor
normalization defined by the inertial tetrad, not necessarily the
present-day density. Equation~\eqref{eq:first_Friedmann_NU} becomes
\begin{equation}
\frac{3H^2}{\kappa^2}=\rho_m+(1+\alpha)m^3
\left(\frac{\dot\phi^2}{2}+V\right),
\label{eq:late_friedmann}
\end{equation}
where $m=1+\alpha(1-a^{-1})$ and $a_\phi=am$.
Since $S_m[h,\Psi_m]$ is independent of the St\"uckelberg fields
when the latter and $h^A{}_{\mu}$ are varied independently, their
first integral \eqref{eq:Stueckelberg_first_integral} remains unchanged.

For $V(\phi)=V_0>0$, define $r_\phi=\dot\phi^2/(2V_0)$ and $N=\ln a$.
The gravitational equations, the St\"uckelberg constraint, and the
scalar-field equation give
\begin{align}
\frac{d\alpha}{dN}
&=3\frac{r_\phi-1}{r_\phi+1}\frac{\alpha(1+\alpha)}{m},
\label{eq:late_alpha_evolution}\\
\frac{dr_\phi}{dN}
&+6\frac{H_\phi}{H}r_\phi
+3\frac{\beta}{m}\frac{d\alpha}{dN}(1-r_\phi)=0,
\label{eq:late_ratio_evolution}
\end{align}
with $\beta=1-a^{-1}$. In the late-time regime $a\gg1$, on the
expanding weak-coupling branch $|\alpha|\ll1$ and $r_\phi\ll1$,
these equations reduce to
\begin{equation}
\frac{d\alpha}{dN}\simeq-3\alpha,\qquad
\frac{dr_\phi}{dN}+6r_\phi\simeq9\alpha.
\label{eq:late_linear_system}
\end{equation}
Together with Eq.~\eqref{eq:Stueckelberg_first_integral}, their asymptotic solution is
\begin{equation}
\alpha\simeq\frac{\mathcal C}{V_0a^3},\qquad
r_\phi\simeq3\alpha,
\label{eq:late_asymptotic}
\end{equation}
up to faster-decaying contributions. Thus potential domination becomes
increasingly accurate; on this nontrivial branch, positivity of the
kinetic energy requires $\mathcal C>0$.

Although $r_\phi\ll1$, its contribution is of the same order as
the leading geometrical correction. Retaining both, the scalar
effective density defined in Eq.~\eqref{eq:rho_eff_cosmo} satisfies
\begin{equation}
\rho_{\rm eff}=(1+\alpha)m^3V_0(1+r_\phi)
\simeq V_0+\frac{7\mathcal C}{a^3},
\label{eq:late_scalar_density}
\end{equation}
so that
\begin{equation}
\frac{3H^2}{\kappa^2}\simeq
V_0+\frac{\rho_{m0}}{a^{3(1+w)}}+\frac{7\mathcal C}{a^3},
\label{eq:late_friedmann_asymptotic}
\end{equation}
to leading order for $a\gg1$, up to relative corrections of
order $a^{-1}$ to the non-universal contribution. The last term
has the background scaling of pressureless matter, although it
originates from the non-universal scalar sector.

For pressureless ordinary matter ($w=0$), define
$D\equiv\rho_{m0}+7\mathcal C>0$. Integrating
Eq.~\eqref{eq:late_friedmann_asymptotic} yields
\begin{equation}
a(t)\simeq\left(\frac{D}{V_0}\right)^{1/3}
\sinh^{2/3}\!\left[
\frac{\sqrt{3\kappa^2V_0}}{2}(t-t_0)
\right].
\label{eq:late_scale_factor}
\end{equation}
For $w=0$, the same asymptotic solution reconstructs the
coupling along the trajectory. Indeed,
$\alpha_{,\phi}^2\simeq3H^2\alpha/(2V_0)$ gives
\begin{equation}
\alpha(\phi)\simeq\frac{\mathcal C}{D}
\sinh^2\!\left[
\kappa\sqrt{\frac{D}{8\mathcal C}}(\phi-\phi_I)
\right].
\label{eq:late_reconstructed_coupling}
\end{equation}
Thus the background solution requires a coupling compatible with
the matter content. Near the de Sitter endpoint, this reduces to
$\alpha\simeq\kappa^2(\phi-\phi_I)^2/8$.

The approximate solution accelerates for
$a>a_{\rm acc}\equiv[D/(2V_0)]^{1/3}$.
Formally, $a\propto(t-t_0)^{2/3}$ near $t_0$, while
$a\propto\exp[\sqrt{\kappa^2V_0/3}\,t]$ at late times.
The former behavior cannot be extrapolated to the early Universe,
since the derivation requires $a\gg1$ and $r_\phi\ll1$.
In particular, $r_\phi(a_{\rm acc})\simeq6\mathcal C/D$;
a controlled description of the acceleration transition requires
$\mathcal C\ll D$ and $a_{\rm acc}\gg1$, implying
$7\mathcal C/\rho_{m0}\ll1$. Thus the present approximation
provides only a subdominant matter-like correction at a controlled
transition, not a dominant dark-matter component. Establishing any
clustering behavior would also require a perturbative analysis.
At late times, $\alpha\to0$, $r_\phi\to0$, and $m\to1$, along with $H^2\to\kappa^2V_0/3$, and therefore, the background approaches
de Sitter expansion.

\section{Concluding remarks}
\label{sec:conclusions}

In this work, we have developed a gauge-invariant teleparallel
framework in which the universality of the gravitational coupling is relaxed for a scalar field, while the gravitational sector retains the TEGR form. The non-universal interaction is encoded in a scalar-sector tetrad whose coupling to the universal translational gauge potential is controlled by a field-dependent gravitational charge. Thus, the construction does not require a modification of the gravitational Lagrangian or the introduction of an additional dynamical gravitational tetrad.

We formulated the theory through a St\"uckelberg completion and
derived the gravitational, scalar-field, and St\"uckelberg equations, together with the antisymmetric constraint and their consistency relations. In particular, the effective energy--momentum tensor sourcing the universal gravitational equations is covariantly conserved, although the scalar energy--momentum tensor is not generally conserved with respect to its effective geometry.

The St\"uckelberg equations are related to the remaining field
equations through gauge symmetry, but their constraints must
be retained after imposing the unitary gauge.
As a first application, we investigated homogeneous and isotropic cosmology. In the kination-dominated regime, the non-universal coupling asymptotically vanishes as the scale factor approaches zero, recovering the standard stiff-fluid evolution. In the potential-dominated regime, we showed that accelerated expansion additionally depends on the spatial deformation induced by the non-universal coupling. In particular, potential domination alone does not guarantee acceleration. We also identified compatible expanding solutions in which departures from universality are dynamically suppressed, including examples approaching power-law inflation or asymptotically de Sitter expansion without imposing the second conventional slow-roll condition.

We further considered a universally coupled perfect fluid
and a constant scalar potential. On the expanding weak-coupling
branch, the non-universal interaction generates an additional
matter-like contribution to the background Friedmann equation,
while the cosmological evolution asymptotically approaches
de Sitter expansion. Within the validity of the approximation,
this contribution remains subdominant and cannot be identified
with a dark-matter component without further investigation.

Future work will focus on cosmological perturbations, the
stability of the resulting solutions, and their observational
consequences for inflation and late-time cosmology.

\acknowledgments
G. Otalora acknowledges Dirección de Investigación,
Postgrado y Transferencia Tecnológica de la Universidad de Tarapacá for financial support through Proyecto
UTA Mayor 4748-26. R. Roy acknowledges financial support from the PhD fellowship of the UTA-ULS-UV consortium. We thank the COSMOGRAV-UTA group for its continuous support and stimulating research environment.

\bibliography{nonuniversal_teleparallel_references}

\end{document}